\documentstyle[11pt,aaspp]{article}
\newcommand{\postscript}[2]
 {\setlength{\epsfxsize}{#2\hsize}
  \centerline{\epsfbox{#1}}}
\def\ref#1{\par\noindent \hangindent=0.4in \hangafter=1 #1 \par}
\def\eqalign#1{\null\,\vcenter{\openup\jot \m@th
  \ialign{\strut\hfill$\displaystyle{##}$&$
     \displaystyle{{}##}$\hfill \crcr#1\crcr}}\,}
\def\tempest%
{\begin{array}{ccc}
1 & 1 & 1 \\
1 & 1 & 1 \\
4 & 3 & 8
\end{array}}

\begin{document}

\title{GALACTIC VERSUS EXTRAGALACTIC}
\title{PIXEL LENSING EVENTS TOWARD M31${}^{1}$}

\author{Cheongho Han}
\author{Andrew Gould$^2$}
\affil{Ohio State University, Department of Astronomy, Columbus, OH 43210}
\affil{cheongho@payne.mps.ohio-state.edu}
\affil{gould@payne.mps.ohio-state.edu}
%lars@helios.ucsc.edu, nsk@astro.washington.edu}
\footnotetext[1]{submitted to {\it Astrophysical Journal}, Preprint:
OSU-TA-3/96}
\footnotetext[2]{Sloan Foundation Fellow}

\begin{abstract}

A new type of gravitational microlensing experiment toward a
field where stars are not resolved is being developed observationally
and theoretically: pixel lensing.
When the experiment is carried out toward the M31 bulge area, events
may be produced both by Massive Compact Halo Objects (MACHOs)
in our Galactic halo and by lenses in M31.
We estimate that $\sim 10-15\%$ of the total events are caused by
Galactic halo MACHOs assuming an all-MACHO halo.
If these Galactic events could be identified, they would provide us
with an important constraint on the shape of the halo.
We test various observables that can be used for the separation
of Galactic halo/M31 events. 
These observables include the Einstein time scale, the 
effective duration of an event, and the flux at the maximum amplification,
but they cannot be used to separate each 
population events.
However, we find that 
most high maximum-flux Galactic halo events can be isolated 
through a satellite-based measurement of the flux difference caused 
by the parallax effect.
For the detection of the flux difference, it is required to
monitor events with an exposure time of 
$\sim 20\ {\rm min}$ by a 0.5 m telescope mounted on a satellite. 
Such observations could be carried out as a minor component
of a mission aimed primarily at
events seen toward the Galactic bulge and  Large Magellanic Cloud.
In addition, proper motion can be used to isolate Galactic halo/M31 events, 
but only for $\sim 5\%$ of high signal-to-noise ratio M31 events and only 
1\% of Galactic halo events.

\end{abstract}

\keywords{
astrometry - dark matter - gravitational lensing - M31
}

\newpage
 
\section{Introduction}
 
Crotts (1992) and Baillon et\ al.\ (1993) have begun a pioneering search
for lensing events of unresolved stars in M31 
by monitoring individual {\it pixels} for the time-dependent
flux induced by one of the many unresolved stars they contain:\ 
`pixel lensing' for short.
For a classical lensing event, the light curve is obtained by subtracting
the reference flux, $B$, from the amplified flux, $F$:
$$
F_{0,i}(A-1)=F-B;\ \ B = F_{0,i} + B_{\rm res},
\eqno(1.1)
$$ where $F_{0,i}$ is the flux of the lensed star before or after the 
event and $B_{\rm res} = \sum_{j\neq i} F_{0,j}$ is the residual 
flux from stars that are not participating in the event.
Here the amplification is
$$
A(x) = {x^2 + 2 \over x(x^2 +4)^{1/2} },\ \
x = [\beta^2 + \omega^2 (t-t_0)^2]^{1/2},
\eqno(1.2)
$$
where $\omega=t_{\rm e}^{-1}$ is the inverse Einstein ring crossing
time, $\beta$ is the dimensionless impact parameter,
and $t_0$ is the time at the maximum amplification.
In an uncrowded field where the blending of stars is not important
($B_{\rm res} \ll F_{0,i}$),
one can directly measure the values of $F$ and $B=F_{0,i}$.
On the other hand, it is impossible to measure the absolute values of
$F$ and $B$ for the case of pixel lensing since stars cannot be resolved.
However, the light curve, $F_{0,i}(A-1)$, can still be obtained 
by subtracting the reference image from those in which a lensing event 
is in progress (Ciardullo et al.\ 1990; Tomaney \& Crotts 1996).

Pixel lensing has several advantages over classical lensing.
First of all, one is not restricted to observing fields where
stars are resolvable.
Second, one can detect ${\cal O} (10^2)$ events
from a season-long observation of M31 in a single field with
moderate resources because a large number of stars are monitored (Han 1996).
Finally, one looks through the Galactic halo when an external galaxy is
monitored for pixel lensing; some fraction of events may be caused
by the Galactic Massive Compact Halo Objects (MACHOs).
If the Galactic MACHOs can be distinguished from events caused by
the lenses in the external galaxy, pixel lensing will considerably
increase the number of Galactic MACHO events.
In addition to the increase in the number of events, one can
use the data toward these additional lines of sight to constrain
the shape of the Galactic halo.
However, events caused by one population are just contaminants
in the study of the other population unless they are separated.
The problem becomes more serious when the fraction of contaminating
events is large.
It is therefore important to determine the fraction of events
caused by each population and to develop methods for distinguishing
them from one another.

We estimate that $\sim 10-15\%$ of total events seen toward the M31 bulge 
area are caused by the Galactic halo MACHOs, assuming an all-MACHO halo.
The first and second year of data
observed by the MACHO group
toward the Large Magellanic Cloud (LMC) gives tentative support to
the hypothesis that MACHOs constitute a significant fraction of
the Galactic halo (D.\ Bennett 1995, private communication).
We test various observables for their ability to distinguish the
Galactic halo events from M31 events.
We find that 
most high maximum-flux Galactic halo events can be isolated 
through a satellite-based measurement of the flux difference caused 
by the parallax effect.
This observational strategy is discussed in \S\ 4.1.
Proper motion can be measured for $\sim 5\%$ of high signal-to-noise 
ratio ($S/N$) M31 events and can also
be used to separate Galactic halo/M31 events.
In addition, 1\% of Galactic events can be isolated by
determining the lower limit of the proper motion, $\mu_{\rm low}$.
Hence, only a small fraction of events can be separated by using 
the proper motion and proper motion-limit measurements.

\section{The Fraction of Galactic Halo Events}
 
\subsection{Optical Depth Estimate}
 
The mean optical depth to an M31 star lensed by the
Galactic halo MACHOs is
$$
\tau_{\rm MW,halo} = {4\pi G \over c^2} \int_{0}^{d_{\rm M31}} dD_{\rm ol}
\rho_{\rm MW,halo} (D_{\rm ol}) D,
\eqno(2.1.1)
$$
where $d_{\rm M31}=770\ {\rm kpc}$ is
the distance to M31, and $D=D_{\rm ol}D_{\rm ls}/D_{\rm os}$, and
$D_{\rm os}$, $D_{\rm ol}$, and $D_{\rm ls}$ are the distances between
the observer, source, and the lens.
For the distribution of matter in the Galactic halo, we adopt a modified
isothermal sphere with a core radius $r_{\rm c,MW}$;
$$
\rho_{\rm MW,halo} (r) =
\cases{
\rho_{0} (r_{\rm c,MW}^2 + R_0^2)/(r_{\rm c,MW}^2 + r^2),
&  $r \le 200\ {\rm kpc}$ \cr
0, & $r > 200\ {\rm kpc}$,\cr}
\eqno(2.1.2)
$$
where $r$ is the distance measured from the Galactic center to a MACHO,
the normalization is
$\rho_{0} = 7.9 \times 10^{-3}\ M_{\odot}\ {\rm pc}^{-3}$,
the core radius is
$r_{\rm c,MW} = 2\ {\rm kpc}$ (Bahcall, Schmidt, \& Soneira 1983), and
$R_0 = 8\ {\rm kpc}$ is the solar galactocentric distance.
The Galactic halo model is the same as that adoted by Griest (1991)
for his computation of the optical depth toward the LMC except that
the halo extends out to $r=200\ {\rm kpc}$, and thus results in higher 
$\tau_{\rm MW,halo}$.
The value of $r_{\rm c,MW}$ is quite uncertain, but it hardly affects
the optical depth determination because M31 is located 
$\sim 119^{\circ}$ away from the Galactic center.
From equations (2.1.1) and (2.1.2), one finds
the optical depth contribution by the Galactic halo MACHOs
to be $\tau_{\rm MW,halo} = 4.4\times 10^{-7}$.

For the M31 halo, we adopt a standard isothermal model with a core radius of 
the form
$$
\rho_{\rm M31,halo}={v^{2}_{\rm c,M31}\over 4\pi G(r^2+r_{\rm c,M31}^{2})}
\eqno(2.1.3)
$$
because it is known that the rotation curve is flat at large radii up to 
$r=30\ {\rm kpc}$ from the M31 center with a rotation velocity of 
$v_{\rm c,M31}=240\ {\rm km}\ {\rm s}^{-1}$ (Kent 1989a).
The core radius is determined from the relation
$$
M_{\rm c,M31} = M_{\rm M31,disk}+M_{\rm M31,bulge}
={v_{\rm c,M31}^{2}\over G}\int_{0}^{\infty} dr r^2 
\left( {1\over r}-{1\over r^2 + r_{\rm c,M31}^{2}} \right),
\eqno(2.1.4)
$$
resulting in $r_{\rm c,M31} = (2/ \pi)(GM_{\rm c,M31}/
v_{\rm c,M31}^2)=6.5\ {\rm kpc}$ with the total mass of the luminous
components taken to be
$M_{\rm c,M31} = 13.2\times 10^{10}\ M_{\odot}$, 
where the total masses of the M31 bulge and disk 
are determined below.
With this density model, the optical depth contributed by 
M31 halo lenses is computed similarly to equation (2.1.1), 
resulting in $\tau_{\rm M31,halo}=1.91\times 10^{-6}$.

On the other hand, the optical depth  along a given line of sight
caused by M31 disk+bulge events is computed by
$$
\tau_{\rm M31} =
{4\pi G \over c^2}
\int_{d_1}^{d_2} dD_{\rm os}\rho_{\rm M31}(D_{\rm os})
\int_{d_1}^{D_{\rm ol}} dD_{\rm ol} \rho_{\rm M31} (D_{\rm ol}) D
\left[ \int_{d_1}^{d_2} dD_{\rm os}\rho_{\rm M31}(D_{\rm os}) \right]^{-1},
\eqno(2.1.5)
$$
because theses populations work both as lenses and sources (self-lensing).
Here $\rho_{\rm M31}=\rho_{\rm M31,disk} + \rho_{\rm M31,bulge}$,
$\rho_{\rm M31,disk}$ and $ \rho_{\rm M31,bulge}$
are the M31 disk and bulge densities, and
$d_1=d_{\rm M31} - 30\ {\rm kpc}$ and $d_2=d_{\rm M31} + 30\ {\rm kpc}$
are the lower and upper boundaries of mass distribution
along the line of sight.
We model the M31 bulge as an oblate spheroid with its 
(unnormalized) matter density as a function of semimajor axis 
provided by Kent (1989b).
The best-fitting value of axis ratio, $c/a=0.75$, is found by fitting
the computed surface brightness (including extinction)
to a $V$-band image of M31 bulge (Han 1996).
In the central region where the M31 bulge mass dominates, the 
rotation velocity is $\sim 275\ {\rm km}\ {\rm s}^{-1}$.
We thus normalize the total M31 bulge mass within $r=4$ kpc by
$M_{\rm M31,bulge}=(275\ {\rm km\ s}^{-1}/v_{\rm c,MW})^{2}M_{\rm MW,bulge}$, 
where $v_{\rm c,MW}=220\ {\rm km\ s}^{-1}$ and 
$M_{\rm MW,bulge}=2\times 10^{10}\ M_{\odot}$ are the adopted rotation speed 
of the Galaxy and Galactic bulge mass within $r=4$ kpc 
(Zhao, Spergel, \& Rich 1995).
This normalization procedure results in 
$M_{\rm M31,bulge}=4\pi\int_{0}^{\infty} dr r^2\rho_{\rm M31,bulge}=
4.9\times 10^{10}\ M_{\odot}$, which is in a reasonable agreement with the 
determined value of $4.0\times 10^{10}\ M_{\odot}$ by Kent (1989b).
The M31 bulge is cut off at $r\sim 8\ {\rm kpc}$.
The M31 disk is modeled by the double exponential disk, i.e., 
$$
\rho_{\rm M31,disk}(R,z) = {\Sigma_0 \over 2h_z}
\exp\left( -{z\over h_z}\right)
\exp\left( -{R\over h_R }\right),
\eqno(2.1.6)
$$
where $\Sigma_{0}=280\ M_{\odot}\ {\rm pc}^{-2}$ is the normalization, and
the radial and vertical scale heights are
$h_{z} = 400\ {\rm pc}$ and $h_R = 6.4\ {\rm kpc}$, respectively 
(Gould 1994c).
This results in 
$M_{\rm M31,disk}(R\le 30\ {\rm kpc})= 
%2\pi\int_{0}^{\infty}dz\int_{0}^{\infty} dR R
%\rho_{\rm M31,disk}= 
8.3\times 10^{10}\ M_{\odot}$.

The resulting total optical depth distribution caused by M31 population 
is shown as a contour map in Figure 1 for various lines of sights.
The projected position $(x',y')$ is defined relative to the major and 
minor axes of the M31 bulge isophotes.
We note that the determined optical depth is subject to
many uncertainties.
The main uncertainty comes from the density distribution of lenses and 
sources.
For example, Braun (1991) has argued that there is no need to invoke a 
massive dark halo component.
Instead of a halo, his model has higher disk and bulge densities, 
c.f., $M_{\rm M31,bulge}= 7.8\times 10^{10}\ M_{\odot}$ and
$M_{\rm M31,disk}= 12.2\times 10^{10}\ M_{\odot}$ within 30 kpc 
from the center.
Pixel lensing experiments may resolve the conflicts in the matter density 
distribution in M31 from the comparison between the theoretical predictions
and future observational result.

\subsection{Event Rate Estimate}
 
Now we estimate the fraction of events
caused by MACHOs in the Galactic halo when a pixel lensing
search is carried out toward the M31 bulge from the ground.
The event rate per angular area $d\Omega$ is determined by
$$
{d\Gamma \over d\Omega}={\Sigma \over \sum_i \phi (F_{0,i})F_{0,i} }
\sum_i \Gamma_0 \beta_{\rm max}(F_{0,i})\phi (F_{0,i});\ \ \
\Gamma_0 = {2 \over \pi} \bar{\omega} \tau,
\eqno(2.2.1)
$$
where $\phi (F_0)$ is the luminosity function (LF)
of the source stars (see \S\ 3)
and $\Sigma$ is the surface brightness in the M31 bulge area
(see Han 1996 for detailed formalism and the surface brightness
distribution).
Here the maximum impact parameter, $\beta_{\rm max}(F_0)$,
within which a lensing event with a source star of luminosity $F_0$ can be
detected for a certain required signal to noise ratio,
$(S/N)_{\rm min}$, is obtained from the equation
$$
{\zeta [\beta_{\rm max} (F_0)] \over \beta_{\rm max} }
= \left( {\Sigma \Omega_{\rm psf} \bar{\omega} \over \pi
F_{0}^{2} \alpha } \right)
(S/N)^2_{\rm min} \eta,
\eqno(2.2.2)
$$
where $\alpha$ is the mean photon detection rate averaged over the
duty cycle of the telescope, $t_{\rm exp}=1\ {\rm day}$ (see below),
$\bar{\omega}=\langle t_{\rm e}\rangle^{-1}$ 
is the inverse value of the median Einstein time scale,
and $\Omega_{\rm psf}\sim \pi \theta_{\rm see}^2 /\ln 4$ is
the angular area of the point spread function (PSF) of the source star 
with a seeing $\theta_{\rm see}$.
The values of $\zeta (\beta )$ are given in
Figure 1 of Gould (1996).
The signal suffers from additional noise due to the sky brightness and
the time variation of the PSF by a factor
$$
\eta^2 = \eta_{\rm sky}^2 + \eta_{\rm psf}^2;
$$
$$
\eta_{\rm sky}^2 = {\Sigma + \Sigma_{\rm sky} \over \Sigma},\ \
\eta_{\rm psf}^2 = {\epsilon_{\bar{m}_{I}}t_{\rm exp} \over 4n_{\rm div} }
\left( \Delta \theta_{\rm see} \over \theta_{\rm see} \right)^2,
\eqno(2.2.3)
$$
where the sky brightness is taken to be
$\Sigma_{\rm sky} \sim 19.5\ {\rm mag}\ {\rm arcsec}^{-2}$
in $I$ band, $\Delta\theta_{\rm see}$ is the seeing difference
between an image and the reference frame,
and $\epsilon_{\bar{m}_I}$ is the rate of photon detection
at the fluctuation magitude of M31, i.e.,
$\bar{m}_I = 23.2$ (Tonry 1991).

The event rate, $\Gamma$, depends on the observational strategy.
Here the event rate is determined under the following observational
conditions.
Event rates for various other strategies are discussed
in detail by Han (1996).
We assume the experiment employs a $2{\rm K}\times 2{\rm K}$ CCD array
with a pixel size of $0.\hskip-2pt ''25$ on a 1-m telescope.
The observation is carried out in $I$ band, and the camera can detect
$12\ {\rm photons}\ {\rm s}^{-1}$ for an $I=20$ star.
The observation is carried out for an average
$4\ {\rm hrs}$ per night during the M31 season,
which is $\sim 1/3$ of a year.
The fractional seeing difference is kept to
$\langle \Delta \theta_{\rm see} /
\theta_{\rm see} \rangle = 5\%$ and each image is obtained by
subdividing the exposure time to prevent saturation of pixels.
In the computation, the sub-integration time is $\sim 20\ {\rm min}$, and
thus $n_{\rm div} = 4\ {\rm hr}/20\ {\rm min} =12$.
The images taken every day are {\it combined} together, and thus $\alpha =
\epsilon \times (4\ {\rm hr}/24\ {\rm hr})\ {\rm photons}\ {\rm hr}^{-1}$,
where $\epsilon$ is the photon detection rate of the camera.

For the computation of event rate, it is required to compute the
median time scale 
$\langle t_{\rm e} \rangle = 1/\bar{\omega}$ [see equation (2.2.2)].
The value of $\langle t_{\rm e} \rangle$ depends on the geometry,
the mass function, and the transverse velocity of the lens system.
The velocity distribution is modeled by a Gaussian with one-dismensional 
dispersion, $\sigma$.
The Galactic MACHO velocity distribution has
$\sigma_{\rm MW,halo}=220/\sqrt{2}\ {\rm km\ s}^{-1}$.
To account for the solar motion around the Galactic center, 
we increase the dispersion by 
$\sigma_{\rm MW,halo}=250/\sqrt{2}\ {\rm km\ s}^{-1}$.
By incresing the Galactic halo lens speed, one can simplfy 
the observer's motion to be $0$.
The M31 halo lens has $\sigma_{\rm M31,halo}=v_{\rm c,M31}/\sqrt{2}
\sim 170\ {\rm km\ s}^{-1}$, while the disk and bulge are modeled by
$\sigma_{\rm M31} = 156\ {\rm km\ s}^{-1}$, based on the measured value
by Lawrei (1983).
With these adopted values of velocity dispersions the transverse 
velocity is determined by
$$
{\bf v} = {\bf v}_{\rm l} - \left[ 
{\bf v}_{\rm s} \left( {D_{\rm ol} \over D_{\rm os}}\right)+ 
{\bf v}_{\rm o} \left( {D_{\rm ls} \over D_{\rm os}} \right)
\right].
\eqno(2.2.4)
$$
where ${\bf v}_{\rm s}$, ${\bf v}_{\rm l}$, and ${\bf v}_{\rm o}$ are 
the velocity of the source, lens, and observer, respectively.
For Galactic halo events, one can approximate 
$D_{\rm ol}/D_{\rm os}\sim 0$ and
$D_{\rm ls}/D_{\rm os} \sim 1$.
Similarily, $D_{\rm ol}/D_{\rm os}\sim 1$ and
$D_{\rm ls}/D_{\rm os} \sim 0$ for M31 halo and M31 disk+bulge 
self-lensing events.
The values of $\sigma$ of the transverse speed
distributions for events of individual populations are listed in Table 1.
The resultant speed distributions of both M31 halo and disk+bulge 
events are similar, so we use a common distribution of 
$\sigma_{\rm M31} = 225\ {\rm km\ s}^{-1}$.

With the transverse speed distribution models,
the time scale distribution of self lensing events (M31 disk+bulge events) 
for a constant mass,
e.g., $1\  M_{\odot}$, is then obtained by
$$
f(t'_{\rm e}) = \int_{d_{1}}^{d_{2}} dD_{\rm os} \rho (D_{\rm os})
\int_{d_{1}}^{D_{\rm ol}} dD_{\rm ol}\rho (D_{\rm ol}) D^{1/2}
\int_0^{\infty} dv
\delta \left( t_{\rm e} - {\sqrt{4GM_{\odot}D}\over cv}\right) vf(v),
\eqno(2.2.5)
$$
where $\delta$ is the Dirac delta function.
For the Galactic and M31 halo events, the time scale distribution can be
approximated by
$$
f(t'_{\rm e}) = \int_{d_{\rm min}}^{d_{\rm max}} dD_{\rm ol} 
\rho (D_{\rm ol}) D^{1/2}
\int_0^{\infty} dv
\delta \left( t_{\rm e} - {\sqrt{4GM_{\odot}D}\over cv}\right) vf(v),
\eqno(2.2.6)
$$
where $d_{\rm min} = 0$ and $d_{\rm max} =200\ {\rm kpc} $ for 
Galactic halo events 
while $d_{\rm min} = d_{\rm M31}-200\ {\rm kpc}$ and 
$d_{\rm max} = d_{\rm M31}$ for M31 halo events.
The additional factors $D^{1/2}$ and $v$ are included to weight events
by their cross section $r_{\rm e} \propto D^{1/2}$ and transverse speed.
Then the true time scale distribution is obtained by convolving
$f(t'_{\rm e})$ with the lens mass function $f_{\rm M}(M)$;
$$
f(t_{\rm e}) = \int dM M^{1/2} f_{\rm M}(M)
\int_0^{\infty} dt'_{\rm e} f(t'_{\rm e})
\delta (t_{\rm e} - M^{1/2}t'_{\rm e}).
\eqno(2.2.7)
$$
Here once again the factor $M^{1/2}$ is included to weight by
$r_{\rm e} \propto M^{1/2}$.
For the mass of Galactic and M31 halo MACHOs, we adopt 
$M=0.1\ M_{\odot}={\rm constant}$,
which is the upper mass limit of hydrogen burning.
Note that $t_{\rm e} \propto M^{1/2}$ for other masses.
However, the majority of M31 disk+bulge events are expected to be caused by 
low mass stars, and thus
we adopt a mass spectrum of local Galactic stars obtained from
HST data (Gould, Bahcall, \& Flynn 1996) for M31 disk+bulge lenses.
The mass spectrum is shown in Figure 2 of Han \& Gould (1996).
To compare the time scale distributions for events from individual 
populations,
we present $f(t_{\rm e})$ for events
at the position
$(x',y')=(1\ {\rm kpc}, 1\ {\rm kpc})$ as representative distributions
in Figure 2.
Note that $f(t_{\rm e})$ for M31 disk+bulge events depends on the 
field of observation; shorter $t_{\rm e}$ toward the center because the 
average separation between lenses and sources 
$(\langle D_{\rm ls}\rangle )$ decreases.
Events from the three populations have very similar distributions 
with $\langle t_{\rm e} \rangle \sim 20\ {\rm days}$
and so the time scale cannot be used as a tool for the separation of
events into different populations.

Finally, the total event rate is then found by
$$
\Gamma = \int_{\Omega_{\rm CCD}} {d\Gamma \over d\Omega} d\Omega,
\eqno(2.2.8)
$$
where the total angular area of the CCD is
$\Omega_{\rm CCD} = (500''\times 500'')$, which is equivalent to 
$\sim (1.87\times1.87)\ {\rm kpc}^2$.
The event rate for each population as a function of $(S/N)_{\rm min}$ 
is shown in Figure 3.
The M31 disk+bulge and M31 halo events have a similar event rates, while
Galactic halo events are $\sim 3-4$ times less frequent
than events from either of the M31 populations, i.e., Galactic 
halo events constitute $\sim 10-15\%$ of the total.

The fraction of Galactic halo events will increase by choosing 
a filed that is located away from the M31 bulge.
As the field is further away from the M31 center, the optical depth for 
the Galactic halo events is virtually the same, but the source star
number, $N_{\ast}$, and thus the total number of events, $N_{\rm tot}$, 
significantly decreases
because $N_{\rm tot} \propto N_{\ast}\tau$.
For example, the number of source stars in the field at
$(1\ {\rm kpc},1\ {\rm kpc})$ is  a factor $\sim 45$ 
larger than that of the field at $(5\ {\rm kpc},5\ {\rm kpc})$. 
Therefore, it would not be a good idea choosing a field too far away
from the M31 bulge just to increase halo/M31 event ratio.

\section{Monte Carlo Simulation}
 
Now we return to the main concern of the article;
what kind of information can one obtain from observables of individual
events?
To answer this question, we carry out a Monte Carlo
simulation of lensing events toward the M31 bulge.
The following are the distributions from which lensing parameters are drawn;
$$
\cases{
g_{\rm M}(M) = \int_0^{M} f(M'){M'}^{1/2}dM', &  \cr
g_{\rm v}(v) = \int_0^{v} f(v')v' dv', &  \cr
g_{\rm r_{\rm e}}(r_{\rm e}) =
\cases{\int dD_{\rm ol} \rho(D_{\rm ol}) D^{1/2} \delta
\left[ r_{\rm e}-\left( {4GM\over c^2} {D_{\rm ol}D_{\rm ls} \over D_{\rm os}}
\right)^{1/2} \right] , &  halo , \cr
\int
dD_{\rm os} \rho(D_{\rm os})
\int
dD_{\rm ol} \rho(D_{\rm ol})D^{1/2}
\delta \left[ r_{\rm e}-\left( {4GM\over c^2}
{D_{\rm ol}D_{\rm ls} \over D_{\rm os}} \right)^{1/2} \right] ,
&  M31, \cr} &  \cr
g_{\beta}(\beta ) = [0,\beta_{\rm max}], &   \cr}
\eqno(3.1)
$$
where $[0,\beta_{\rm max}]$ represents a uniform distribution in the range
$0 \leq \beta \leq \beta_{\rm max}$.

The source star LF is based primarily on the LF of
Galactic bulge stars determined by J.\ Frogel (1995, private communication).
However, this determination is incomplete at the faint end
($M_I \ge 3.6$).
If it were adopted without correction, the fraction of luminous
stars would be overestimated.
The net effect would then be the overestimation of the event rate
because events are more likely to be detected for luminous stars.
We model the faint part of the LF by adopting the LF of stars in the
solar neighborhood determined by Wielen, Jahreiss, \& Kr\"uger
(1983) for $3.6 < M_I < 8$ and by
Gould, Bahcall, \& Flynn  (1996) for $8 < M_I < 14$.
The corrected model LF yields a fluctuation magnitude
of $\sum_i \phi(F_{0,i})F_{0,i}^2 /\sum_i \phi(F_{0,i})F_{0,i}=23.5$,
which matches well with Tonry's (1991)
determination of $\bar{m}_{I}=24.3$ before the extinction correction.
The fraction of light from the stars in the corrected part of the LF
is $14.7\%$, which is not negligible.
The LFs before ({\it thin} line) and after the model correction 
({\it thick} line) are shown in the first panel of Figure 5.
For each luminosity, the value of
$\beta_{\rm max}$ is obtained from equation (2.2.2).

Detectable events are those that satisfy the condition
$\beta \le \beta_{\rm max}$.
In addition, events must satisfy the following two conditions.
The first one is that the effective duration of an event, 
$t_{\rm eff}=\beta t_{\rm e}$, must be long enough
for the reasonable construction of light curve,
i.e., $t_{\rm eff} \ge 0.5\ {\rm day}$.
Second, the angular size of the source star must be smaller than
the angullar Einstein ring size, i.e.,
$\theta_{\ast}/\theta_{\rm e} \le \beta_{\rm max}$, to avoid the
maximum flux suppression due to the finite size source effect
(Nemiroff \& Wickramashinghe 1994; Witt \& Mao 1994;
Gould 1994a; Simmons, Newsam, \& Willis 1995).
Both Galactic and M31 halo events have very similar physical parameters.
They have same masses ($0.1\ M_{\odot}$) and similar transverse speeds
($177\ {\rm km\ s}^{-1}$ versus $225\ {\rm km\ s}^{-1}$).
In addition, due to the equivalence between $D_{\rm ol}$ for
Galactic lenses to $D_{\rm ls}$  of M31 halo events, vice versa,
both population events have very similar time scales, as already shown
in Figure 2, and Einstein ring sizes, $r_{\rm e}\propto \sqrt{MD}/v$.

\section{Separation of Halo/M31 Events}
 
Separating Galactic halo/M31 events is important
for the better understanding of
the global properties of the Galactic halo such as its flattening.
Because Galactic halo events account for only $\sim 10-15\%$ of
the total, it is possible to use observations toward M31 to
constrain the Galactic halo only if the halo events can be identified.
Sackett \& Gould (1993) have proposed that one can constrain
the shape of the Galactic halo by studying the optical depths
toward LMC and Small Magellanic Cloud (SMC).
To understand the role of M31 in determining the shape of the halo,
we write the Galactic coordinates of the SMC, LMC, and M31 as
$(l,b)=(-57^{\circ},-44^{\circ})$,
$(-80^{\circ},-33^{\circ})$, and $(121^{\circ},-21^{\circ})$.
Thus, the lines of sight toward SMC, LMC, and M31 are progressively
closer to the Galactic plane and further from the Galactic center.
This implies that by adding the line of sight to M31,
one increases the leverage on the shape of the Galactic halo
relative to observing the LMC and SMC alone.

Additionally, halo events are contaminants
for the study of MACHOs in M31.
If the Galactic halo is flattened (Dubinski \& Calberg 1991;
Katz 1991; Sackett \& Sparke 1990), the contamination of events toward M31
by the Galactic halo events would be more significant
than our estimate.
If one can separate a {\it large} fraction of M31 events from the
Galactic halo events, a better study about the nature of M31 MACHOs
will be possible.
In this section, we investigate the methods and required observables
that can be used to isolate the two populations.

We begin with a brief discussion of the general observables in pixel lensing.
Near the peak of a light curve, the flux is approximated by
$$
F\sim {F_{\rm max} \over [1+(t/t_{\rm eff})^2]^{1/2} };\ \
F_{\rm max} = {F_0 \over \beta},\ \ t_{\rm eff} = \beta t_{\rm e},
\eqno(4.1)
$$
since $A\rightarrow 1/x$ and $x = [\beta^2+(t/t_{\rm e})]^{1/2}$.
Here $F_{\rm max}$ is the flux at the maximum amplification.
The values of $F_{\rm max}$ and $t_{\rm eff}$ are {\it directly} 
measurable for individual moderately high amplification events.
The distributions of $t_{\rm eff}$ and $F_{\rm max}$ are
determined for events with $(S/N)_{\rm min}=20$ and
$(S/N)_{\rm min} = 70$ by the Monte Carlo
simulation and are shown in Figure 4.
Since we already know that Galactic and M31 halo events have similar 
physical paramaters, and so similar for $F_{\rm max}$ and $t_{\rm eff}$, 
we show only distributons of Galactic halo and M31 disk+bulge events.
We present their distributions at the representative positions
of the M31 bulge area at $(x',y')=(0.5\ {\rm kpc}, 0.5\ {\rm kpc})$
({\it thick} line) and $(1.5\ {\rm kpc}, 1.5\ {\rm kpc})$ ({\it thin} line).
The distributions for the Galactic (and M31) halo and M31 
disk+bulge self-lensing events are
similar to each other, and thus neither quantity can be directly used
for the separation  of individual populations.
Although $t_{\rm e}$ can be measured for some of the high
$S/N$ ($\ge 70$) events, it is not in general a direct observable.
However, one can infer $t_{\rm e}$ from $t_{\rm eff}$ and
$F_{\rm max}$ provided that the color of the source star is known.
The color of a source star can still be measured
from the maximum flux since $F_{{\rm max},I}/F_{{\rm max},V}
= F_{0,I}/F_{0,V}$ and the event is achromatic.
The approximate value of the luminosity, $L$, can then be determined
using a color-mag relation.
However, there can be different types of stars with the same color, e.g., 
lower main sequence stars and giants.
Therefore, there will be a two-fold degeneracy in determining $L$ 
for low $S/N$ events. 
For high $S/N$ events, 
$t_{\rm e}$ (and so $F_{0}$) can be approximately determined 
from the light curve, thereby breaking the degeneracy and 
allowing unambiguous determination of $L$ from the color.
Fortunately, the stellar population expected to be detected at low 
$S/N$ are nearly all main sequence stars, so it should be possible to 
determine $t_{\rm e}$ in nearly all cases.
In Figure 5, we plot the LF of source stars (before amplification)
for the expected events with different values of threshold $S/N$;
$(S/N)_{\rm min}=10$ and $70$ in the second and third panels, respectively.
At lower $(S/N)_{\rm min}$, most detectable events are from 
main-sequence stars.
As the $(S/N)_{\rm min}$ increases, a significant fraction of events 
comes from upper main-sequence and clump giant stars.
As shown in Figure 2,  however, $t_{\rm e}$ is not useful for
separating halo/M31 events, either.
Two other quantities can be measured under special circumstances:
the displacement of the source star in the Einstein ring due to
parallax effect and the proper motion of the lens, $\mu$.

%However, there can be different types of stars with the same color, e.g.,
%lower main sequence and giants.
%Fortunately, the stellar population expected to be detected is
%a nearly unique function of $(S/N)_{\rm min}$.
%On the other hand, the lower main-sequence contribution to the events is
%5not significant for high $S/N$ events.
%Therefore, the luminosity of the source star can be determined without
%serious ambiguity.
%To the extent that $L$ (and so $F_0$) can be determined,
%$\beta(=F_0/F_{\rm max})$ and thus $t_{\rm e}(=t_{\rm eff}/\beta)$
%can also be determined.
%We also note that stars fainter than $I\sim 30$ make a negligible
%contribution to events, and thus the uncertainty in the detailed structure
%of the model LF does not significantly affect the result.

\subsection{Parallax Measurement}
 
One of the tools for the separation of both Galactic halo/M31 events 
is provided by the parallax effect.
The principle is simple.
By observing an event at two different positions,
one from the ground and the other from a heliocentric satellite,
the light curves display a
displacement in the Einstein ring, $\Delta x$
(Refsdal 1966; Gould 1994b).
Then the flux measured from the ground will be different from that
measured from the satellite for the Galactic halo events
because the ground-based photometry is carried out when
the event is near the maximum amplification, while the flux measured
from the satellite, $F_{\rm sat}$, is not.
However, the flux difference due to parallax effect 
will be extremely small for M31 self-lensing events due to their negligible
amount of parallax $[\Delta x \sim (D_{\rm ls}/D_{\rm os})
({\rm AU}/r_{\rm e}) \sim 10^{-2}]$.
Therefore, when the difference between the fluxes, $\Delta F$,
from ground and satellite-based photometry is detected, the event
is almost certainly caused by a Galactic halo MACHO.
It will be difficult to precisely measure $\Delta x$;
the signal-to-noise ratio will be low and a typical event lasts
only $\langle t_{\rm eff}\rangle \sim 1\ {\rm day}$, which is
too short for the construction of detailed light curve.
However, measurement of $\Delta F$ will still be feasible, and
this by itself allows the Galactic events to be recognized.

%Let us assume a reasonable observational environment for the 
%measurement of $\Delta F$.
%The expected number of events in process at a given moment is
%$\sim N_{\rm tot} \times (\langle t_{\rm eff}\rangle / t_{\rm obs})$,
%where $N_{\rm tot} \sim 100\ {\rm events}$ is the total number
%of events that are expected to be detected at
%$(S/N)_{\rm min}=10$ during an M31 season of
%$t_{\rm obs} = 4\ {\rm month}$, and the typical effective
%time scale is $\langle t_{\rm eff} \rangle \sim 1\ {\rm day}$.
%Thus $\sim 1\ {\rm event}$ on average is expected to be at or near the maximum
%magnification seen from the ground.
%Note that unlike classical lensing events, pixel lensing events
%can only be noticed during $t_{\rm eff}$.
%Therefore, photometry with a frequency $\sim 1-2\ {\rm day}^{-1}$
%is required.
%Since the photometry can be done with a relatively short exposure
%time, $\Delta t\sim 20\ {\rm min}$ (see below), it
%can be carried out as a part of a larger project whose principal
%mission is parallax measurements of events toward the Galactic
%bulge and LMC.
%The alternative method would be the daily photometry over the whole
%M31 bulge area.
%However, this method would require a large CCD to cover a large angular
%area and one would have to solve the problem of transmitting
%large quanties of data to a ground station.

Now, we answer the question: what fraction of Galactic halo events
can be distinguished in this way?
For this determination, we assume that the satellite photometry will be
carried out with a camera on a 0.5 m telescope mounted on a 
heliocentric satellite. 
The CCD camera is exposed for $\Delta t = 20\ {\rm min}$ and can detect
$3\ {\rm photons}\ {\rm s}^{-1}$ for an $I=20$ mag star.
We assume that the seeing is slightly larger than the diffraction limit:
$\theta_{\rm see} = 0''\hskip-2pt .5$.
The background surface brightness varies in the range
$\Sigma_{I} \sim 19-22\ {\rm mag}\ {\rm arcsec}^{-2}$ in most of
the M31 bulge area except for the part very close to the central region.
%To be conservative, we adopt $\Sigma_{I}=19\ {\rm mag}\ {\rm arcsec}^{-2}$.
The read-out noise is assumed to be negligible.
Then the signal to noise ratio from the satellite observation is 
computed by
$$
(S/N)_{\delta x} = {\Delta F \over (F_{\rm sat} + F_{\rm back})^{1/2}}
\Delta t^{1/2},
\eqno(4.1.1)
$$
where $F_{\rm sat}$ is the measured flux from the satellite,
the background flux is
$F_{\rm back} = F_{I,0} 10^{-0.4\Sigma_I}\Omega_{\rm psf}$,
and $F_{I,0}=3\times 10^8\ {\rm photons}\ {\rm s}^{-1}$ is the
flux for $I=0$.

The flux difference due to parallax is computed as follows.
The Earth-satellite separation vector projected
onto the lens plane is
$\Delta {\bf r} = {\bf R} -
({\bf R}\cdot \hat{\bf s})\hat{\bf s},$
where  ${\bf R}$ is the 3-dimensional Earth-satellite vector and
$\hat{\bf s}$ is the unit vector toward M31 from the Sun.
In scalar form, the separation is given by
$$
\Delta r = R (1-\cos^2\beta_{\rm ec}\cos^2 \psi )^{1/2},
\eqno(4.1.2)
$$
where $\beta_{\rm ec} = 33^{\circ}\hskip-2pt .3$ is the ecliptic
latitude of M31 and $\psi$ is the phase of the orbit (Gould 1995).
During an M31 season,
$\psi$ varies in the range $60^{\circ} \leq \psi \leq 180^{\circ}$.
The amount of displacement in Einstein ring due to the parallax is
then calculated by
$$
\delta x = \Delta x \cos \Phi =
{\Delta r \over \tilde{r}_{\rm e}} \cos \Phi;\ \
\cos \Phi = \hat{\bf r}\cdot \hat{\bf x},
\eqno(4.1.3)
$$
where the projected Einstein radius is
$\tilde{r}_{\rm e} = r_{\rm e}(D_{\rm os}/D_{\rm ls})$.
For Galactic halo events, $\tilde{r}_{\rm e} \sim r_{\rm e}$ since
$D_{\rm os}/D_{\rm ls} \rightarrow 1$.
The angle $\Phi$ varies in the range $[0,2\pi]$, i.e., random orientation,
and the distribution of $r_{\rm e}$ is obtained from equation (3.1).
Then the flux difference is computed by
$$
\Delta F = F_{\rm max} - F_{\rm sat}
=\left[ A(\beta )-A(\beta + \delta x) \right] F_0.
\eqno(4.1.4)
$$

In Table 2, we present the numbers, $N_{\rm sep}$, and fractions, 
$N_{\rm sep}/N_{\rm MW,halo}$,
of Galactic halo events that can be separated from M31 events by
measuring $\Delta F$ for various values of $F_{\rm max}$.
Both values are determined under the criteria of
$(S/N)_{\delta x}\ge 3$ and $5$ (i.e., 3 and 5 $\sigma$ levels).
The numbers of events due to the Galactic halo, $N_{\rm MW,halo}$, 
and due to all populations (M31 halo, disk+bulge, and Galactic halo),
$N_{\rm tot}$, are those that can be detected with $S/N \ge 20$ from 
ground observations.
For event with $F_{\rm max} \le 23\ {\rm mag}$, nearly all
Galactic halo events can be separated, while the fraction decreases
significantly for low $F_{\rm max}$ events because the stars are 
too faint to be seen.
(To avoid confusion, we note that these faint events can still 
be detected from the ground because the image combination process yields 
a higher $S/N$.)
Therefore, the best target for $\Delta F$ measurement would be 
high $F_{\rm max}$ events.
Since the photometry can be done with a relatively short exposure
time, $\Delta t\sim 20\ {\rm min}$, it
can be carried out as a part of a larger project whose principal
mission is parallax measurements of events toward the Galactic
bulge and LMC.

\subsection{Proper Motion}
 
Proper motion of MACHOs can also be used for the separation of
Galactic halo/M31 events.
For a classical lensing event in which a MACHO passes over or very close
to the face of a star, the part of the star closer to the lens is
amplified more than other parts because the source star is not
a perfect point source: differential amplification.
By measuring the amount of deviation of the light curve from that of
point source, one can
measure the ratio
$$
x_{\ast} = {\theta_{\ast}\over \theta_{\rm e}},
\eqno(4.2.1)
$$
where $\theta_{\ast} = R_{\ast}/D_{\rm os}$ is the angular size of
the source.
Since $\theta_{\ast}$ can be determined from the measured color
(and Stefan's law), one can uniquely determine the
$\theta_{\rm e}$ and the proper motion of the MACHO,
$\mu = \theta_{\rm e} / t_{\rm e}$ (Gould 1994a).
For pixel lensing events in which the lens transits the source,
it is also possible to measure the proper motion.
However, the process is somewhat different from that in classical lensing.
The source cannot be resolved, so it is impossible to determine
$\theta_{\ast}$ from Stefan's law.
Instead, one would estimate $R_{\ast}$ from the observed color
and the color-mag relation, and then compute $\theta_{\ast}$.
Generally, it is also not possible to measure $t_{\rm e}$.
However, from the deviation of light curve from its point source form,
one can determine $t_{\ast}$, which is the time it takes for the lens
to cross the stellar radius.
The proper motion is then given by $\mu = \theta_{\ast}/t_{\ast}$.
Once $\mu$ is measured, one can easily
distinguish halo/M31 events
because the values of $\mu$ are quite different
for the Galactic halo and M31 events.
However, the fraction of events for which proper motions can be
measured is small.
For M31 disk+bulge self-lensing events with $S/N \ge 50$, the fraction is
$\sim 5\%$ and similar to M31 halo events.
Since the fraction scales inversely with $\theta_{\rm e}$,
the Galactic fraction is smaller by a factor $\sim 10^{-2}$,
and is therefore completely negligible.
 
In the more usual case when the lens does not transit the source star
$[\beta > (\theta_{\ast}/\theta_{\rm e})$], 
one can still set the lower limit of the proper motion of
individual events by
$$
\mu > \mu_{\rm low} = {\theta_{\ast} \over t_{\rm eff} }.
\eqno(4.2.2)
$$
Then a Galactic halo event can be isolated whenever
$\mu_{\rm low}$ is too high to be consistent with M31 self-lensing.
We determine the fraction of Galactic halo events that can be isolated
in this way by setting the upper limit of the M31
lens population at
$1500\ {\rm km}\ {\rm s}^{-1}d_{\rm M31}^{-1}$.
Unfortunately, only $\sim 1\%$ of
Galactic halo events can be isolated using proper motion lower limits.

\acknowledgments
This work was supported by a grant AST 94-20746 from the NSF.

\vfill\eject

\centerline{\bf FIGURE CAPTION}
\bigskip

\noindent
{\bf Figure 1}:
The optical depth distribution toward the M31 bulge.
The contours are drawn in units of $10^{-6}$.
In the computation, the uniform contribution by the M31 halo,
$\tau_{\rm M31,halo}=1.91\times 10^{-6}$, is included.
The total optical depth both by M31 and Galactic halo lenses
is obtained by adding $\tau_{\rm halo}$ to the 
values marked on the contour, where $\tau_{\rm halo}=4.4\times 10^{-7}$.
With the adopted distance to M31 of $d_{\rm M31}=770\ {\rm kpc}$, 
1 arcsec corresponds to $\sim 3.73\ {\rm pc}$.

\noindent
{\bf Figure 2}: 
The Einstein time scale distributions, $f(t_{\rm e})$, for the 
Galactic halo, M31 self-lensing, and M31 halo events at the
position $(x',y') = (1\ {\rm kpc}, 1\ {\rm kpc})$.
The time scale distributions for individual populations are 
very similar one another.

\noindent
{\bf Figure 3}:
The event rates, $\Gamma$, for individual population
events as a function of the threshold signal-to-noise ratio,
$(S/N)_{\rm min}$.
The observational conditions are described in \S\ 2.2.

\noindent
{\bf Figure 4}:
The effective time scale and maximum flux distributions for
$(S/N)_{\rm min}=20$ and $70$ for the Galactic halo ({\it dotted} lines)
and M31 disk+bulge ({\it solid} lines) events.
Because the distributions are location dependent, those
at $(x',y')=(0.5\ {\rm kpc}, 0.5\ {\rm kpc})$ ({\it thick} lines)
and $(1.5\ {\rm kpc},1.5\ {\rm kpc})$ ({\it thin} lines) are presented
as representative distributions.

\noindent
{\bf Figure 5}:
The LFs of detectable events with different values of $(S/N)_{\rm min}$.
The original LF, modeled by combining the Galactic bulge LF
(J.\ Frogel, private communication), {\it thin} line, and the LFs of 
Gould et al.\ (1996) and Wielen et al.\ (1983), {\it thick} line, is 
shown in the first panel.
The (unamplified) LF of detectable events with 
$(S/N)_{\rm min}=10,\ {\rm and}\ 70$
are shown in the second and third panels.

\vfill\eject
 
%Table
\bigskip
\begin{center}
\centerline{\small TABLE 1}
\smallskip
\centerline{\small TRANSVERSE VELOCITY}
\smallskip
\begin{tabular}{lr}
\hline
\hline
\multicolumn{1}{c}{event} &
\multicolumn{1}{c}{dispersion\ ($\sigma^2$)} \\
\multicolumn{1}{c}{population} &
\multicolumn{1}{c}{$({\rm km\ s}^{-1})^2$} \\
\hline
Galactic halo events &  $177^2 + 0^2 = 177^2$ \\
M31 halo events &  $170^2 + 156^2 = 230^2$ \\
M31 (disk+bulge) events  & $156^2 + 156^6 = 221^2$ \\
\hline
\end{tabular}
\end{center}
\smallskip
\noindent
{\footnotesize
\qquad NOTE.---
The transverse speed distribution for each population event.
Each speed has a Gaussian distriburion with a dispersion
listed.
}
\bigskip

\newpage
%Table
\bigskip
\begin{center}
\centerline{\small TABLE 2}
\smallskip
\centerline{\small GALACTIC HALO EVENTS SEPARABLE BY PARALLAX EFFECT}
\smallskip
\begin{tabular}{ccccccc}
\hline
\hline
\multicolumn{1}{c}{events} &
\multicolumn{1}{c}{$N_{\rm tot}$} &
\multicolumn{1}{c}{$N_{\rm MW,halo}$} &
\multicolumn{2}{c}{$(S/N)_{\delta x} \ge 3$} &
\multicolumn{2}{c}{$(S/N)_{\delta x} \ge 5$} \\
\multicolumn{1}{c}{with $F_{\rm max}$} &
\multicolumn{1}{c}{} &
\multicolumn{1}{c}{} &
\multicolumn{1}{c}{$N_{\rm sep}$} &
\multicolumn{1}{c}{$N_{\rm sep}/N_{\rm MW,halo}$} &
\multicolumn{1}{c}{$N_{\rm sep}$} &
\multicolumn{1}{c}{$N_{\rm sep}/N_{\rm MW,halo}$} \\
\hline
$\le 21\ {\rm mag}$ & \ 2.62 & 0.13 & 0.13 & 100\% & 0.13 & 100\% \\ 
$\le 22\ {\rm mag}$ & \ 7.42 & 0.37 & 0.36 & \ 98\% & 0.36 & \ 97\% \\ 
$\le 23\ {\rm mag}$ & 18.95 & 0.96 & 0.92 & \ 96\% & 0.84 & \ 87\% \\ 
$\le 24\ {\rm mag}$ & 41.59 & 2.38 & 1.36 & \ 57\% & 0.86 & \ 36\% \\ 
$\le 25\ {\rm mag}$ & 65.31 & 5.19 & 1.40 & \ 27\% & 0.94 & \ 18\% \\ 
total               & 95.00 & 10.0 & 1.70 & \ 17\% & 1.20 & \ 12\% \\ 
\hline
\end{tabular}
\end{center}
\smallskip
\noindent
{\footnotesize
\qquad NOTE.---
The numbers, $N_{\rm sep}$, and fractions, $N_{\rm sep}/N_{\rm MW,halo}$, 
of Galactic halo events that can be separated from M31 events by 
measuring $\Delta F$ for various values of $F_{\rm max}$.
Both values are determined under the criteria of 
$(S/N)_{\delta x}\ge 3$ and $5$ (i.e., 3 and 5 $\sigma$ levels).
The numbers of events due to the Galactic halo, $N_{\rm MW,halo}$,
and due to all populations (M31 halo, disk+bulge, and Galactic halo),
$N_{\rm tot}$, are those that can be detected with $S/N \ge 20$ from
ground observations.
For event with $F_{\rm max} \le 23\ {\rm mag}$, nearly all 
Galactic halo events can be separated, while the fraction decreases 
significantly for low $F_{\rm max}$ events.
}
\bigskip

\newpage
\begin{figure}
\postscript{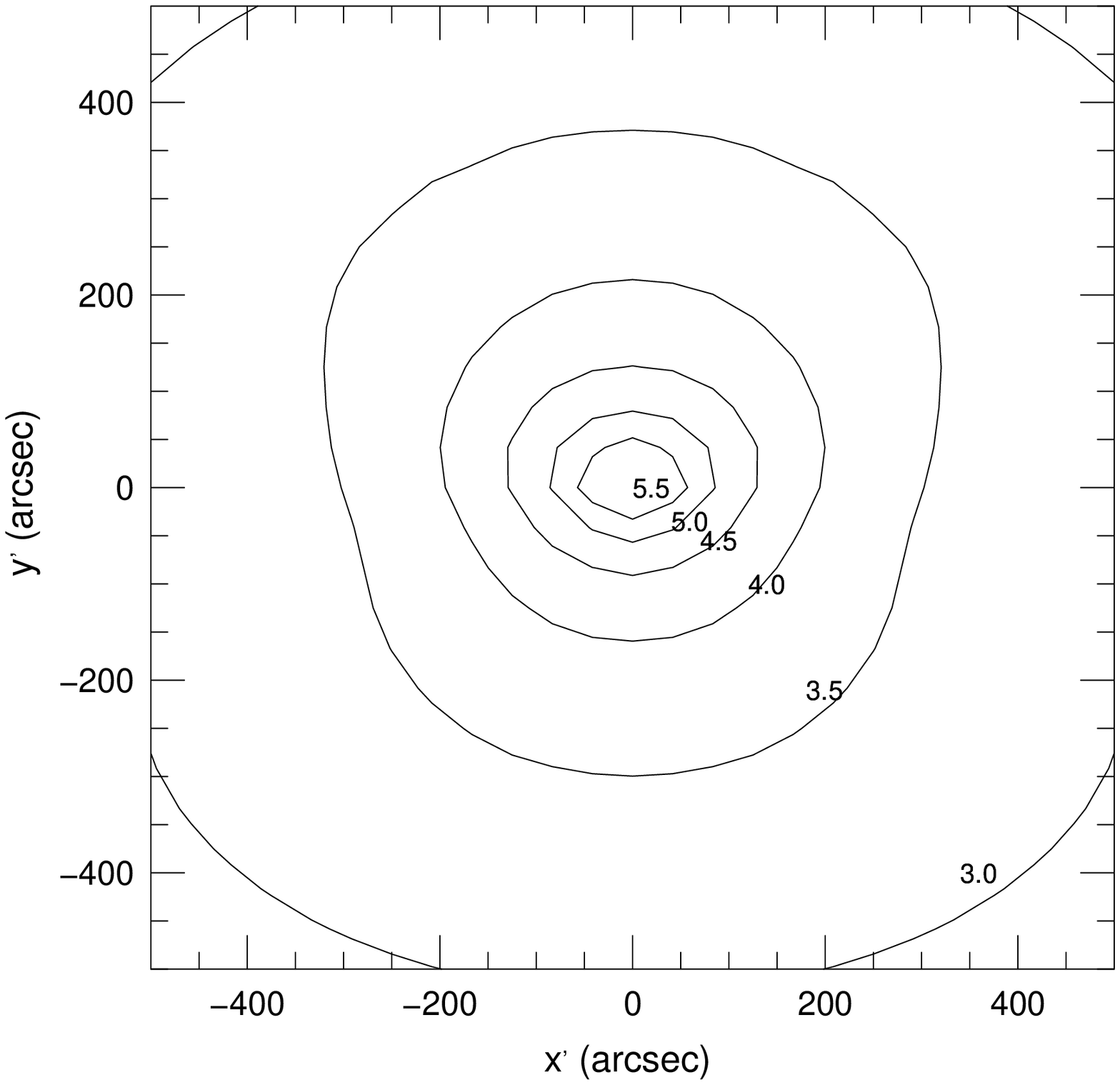}{1.1}
\end{figure}
 
\newpage
\begin{figure}
\postscript{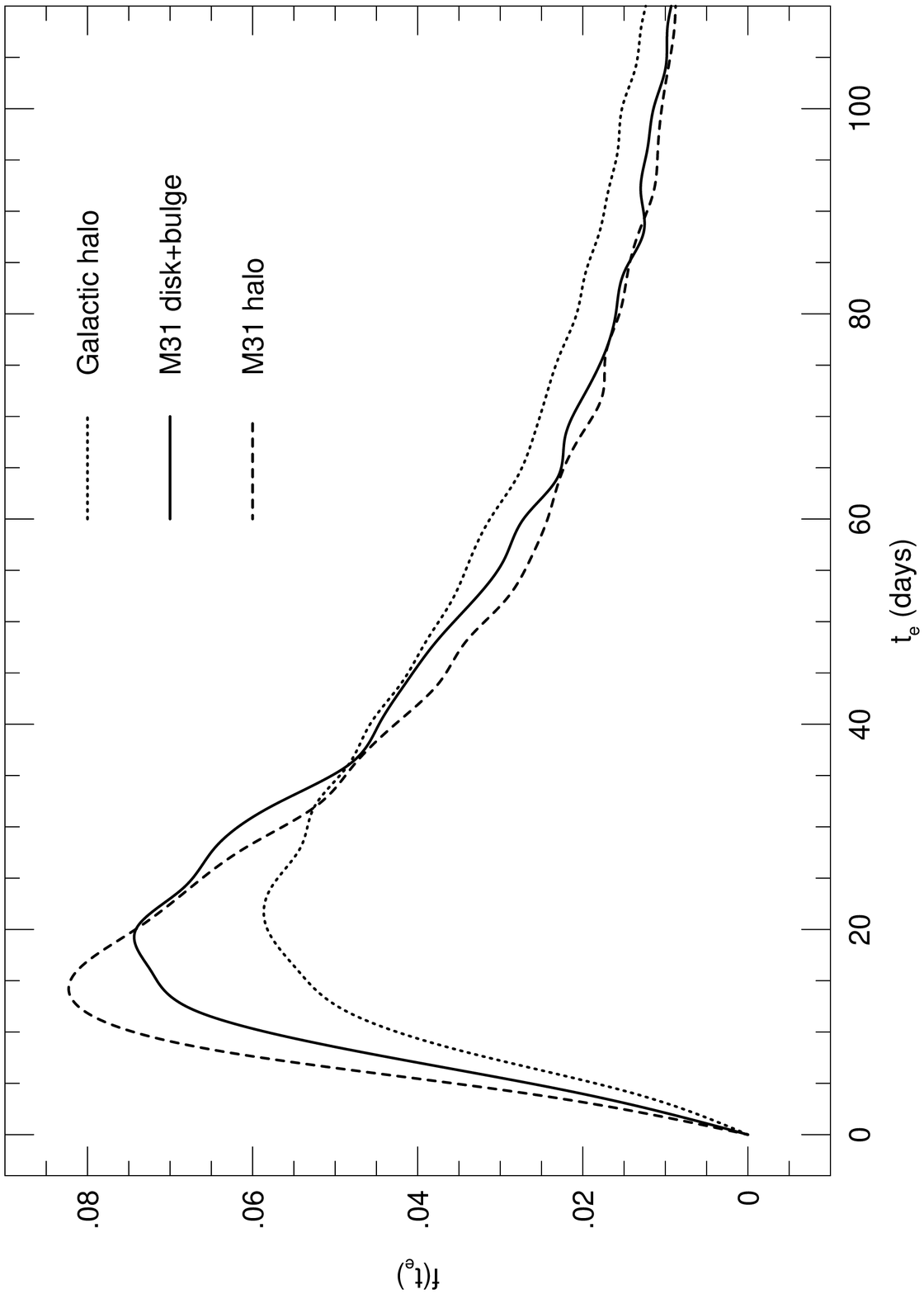}{1.1}
\end{figure}
 
\newpage
\begin{figure}
\postscript{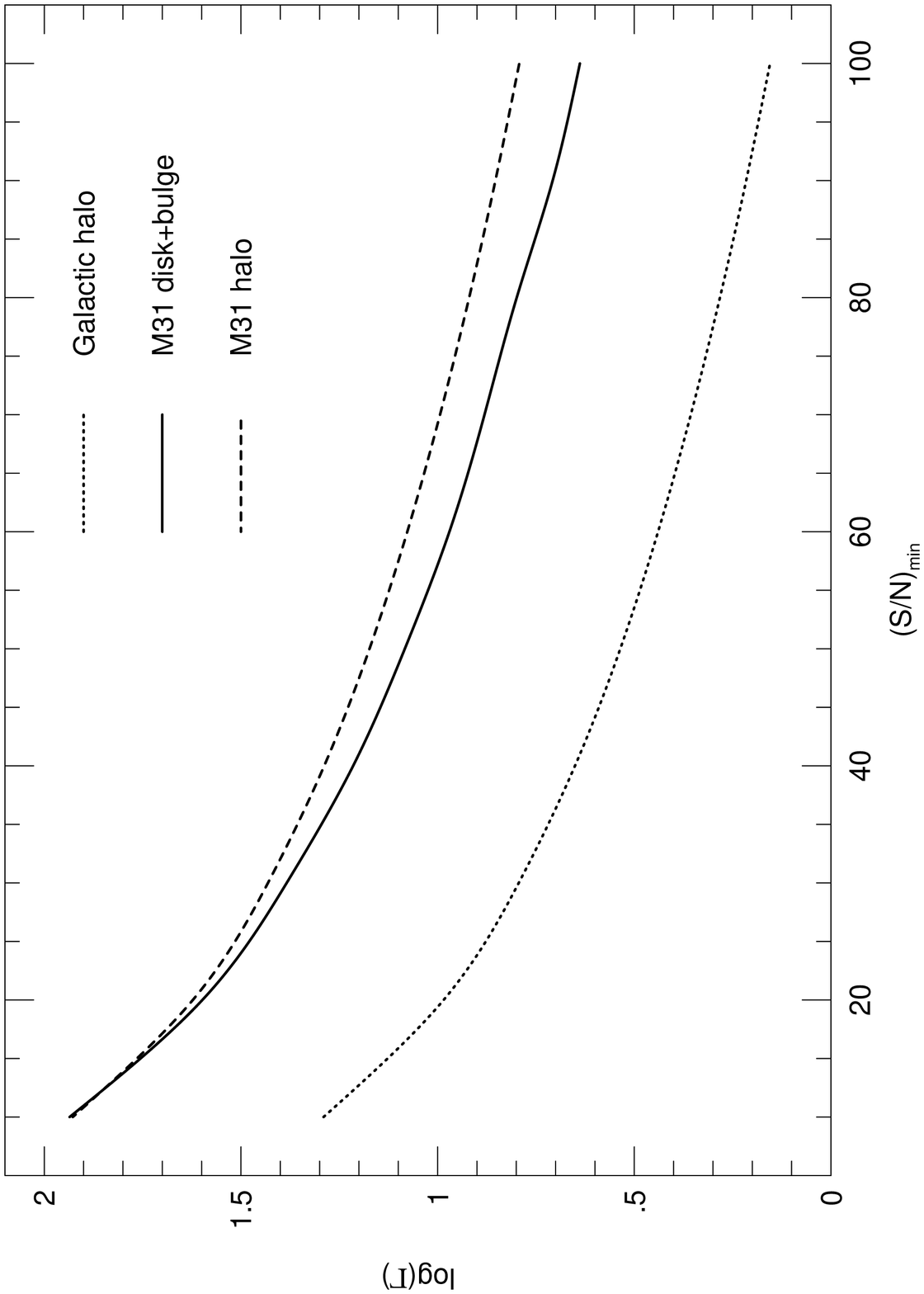}{1.1}
\end{figure}
 
\newpage
\begin{figure}
\postscript{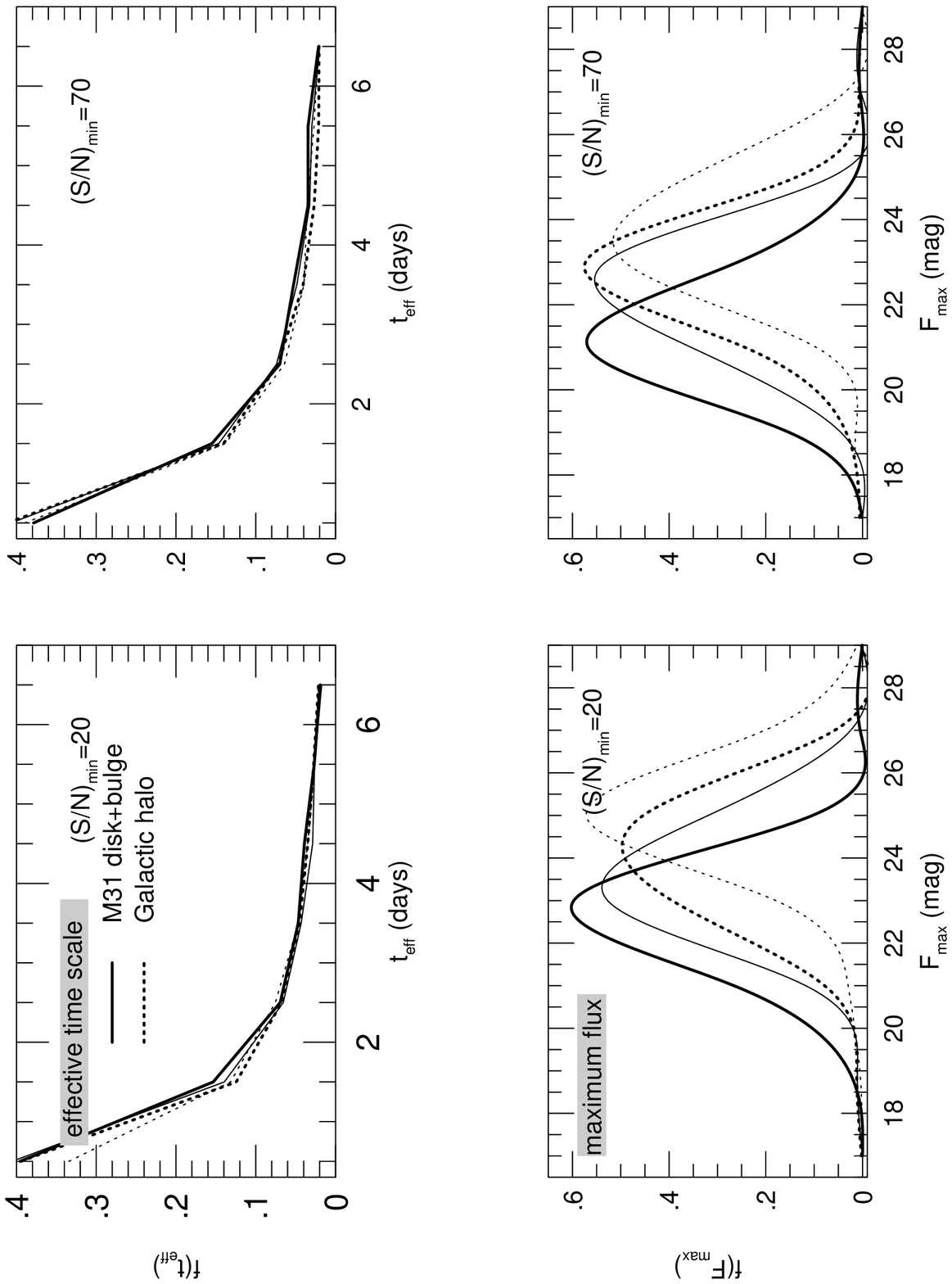}{1.1}
\label{fig4}
\end{figure}
 
\newpage
\begin{figure}
\postscript{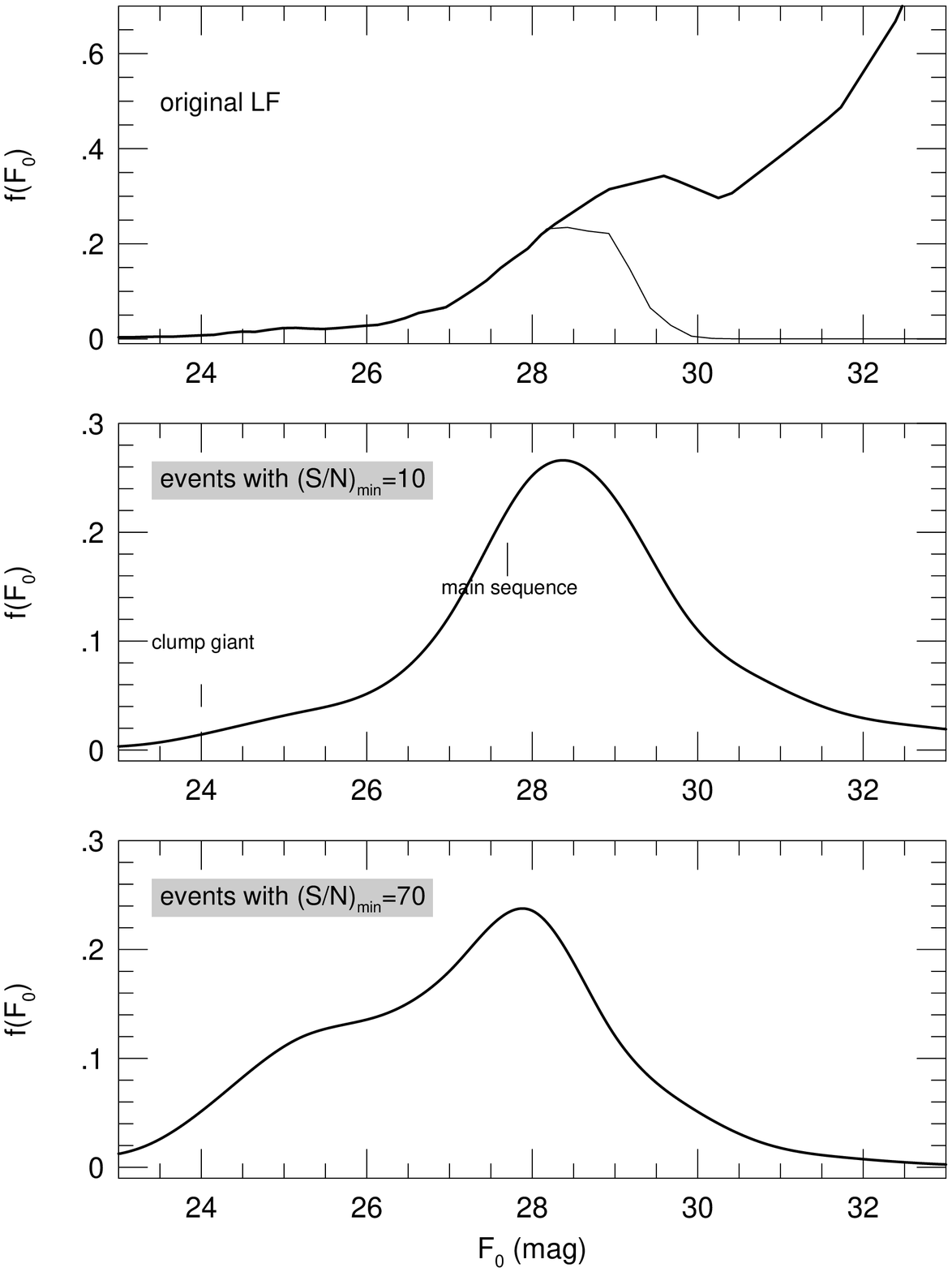}{1.1}
\end{figure}

\end{document}